\documentclass[pdflatex,sn-nature]{sn-jnl}

\usepackage{graphicx}%
\usepackage{multirow}%
\usepackage{amsmath,amssymb,amsfonts}%
\usepackage{amsthm}%
\usepackage{mathrsfs}%
\usepackage[title]{appendix}%
\usepackage{xcolor}%
\usepackage{textcomp}%
\usepackage{manyfoot}%
\usepackage{booktabs}%
\usepackage{algorithm}%
\usepackage{algorithmicx}%
\usepackage{algpseudocode}%
\usepackage{listings}%
\usepackage{lineno}

\begin{document}

\title[Article title]{Satellite monitoring uncovers progress but large disparities in doubling crop yields}  


\author[1,2,3]{\fnm{Katie} \sur{Fankhauser}}\email{katie.fankhauser@colorado.edu}

\author[1]{\fnm{Evan} \sur{Thomas}}\email{evan.thomas@colorado.edu}

\author*[1,2,3]{\fnm{Zia} \sur{Mehrabi}}\email{zia.mehrabi@colorado.edu}
\affil[1]{\orgdiv{Mortenson Center in Global Engineering \& Resilience}, \orgname{University of Colorado Boulder}, \orgaddress{\city{Boulder}, \state{CO}, \country{USA}}}

\affil[2]{\orgdiv{Better Planet Lab}, \orgname{University of Colorado Boulder}, \orgaddress{\city{Boulder}, \state{CO}, \country{USA}}}

\affil[3]{\orgdiv{Department of Environmental Studies}, \orgaddress{\city{Boulder}, \state{CO}, \country{USA}}}


\abstract{

High-resolution satellite-based crop yield mapping offers enormous promise for monitoring progress towards the SDGs. Across 15,000 villages in Rwanda we uncover areas that are on and off track to double productivity by 2030. This machine learning enabled analysis is used to design spatially explicit productivity targets that, if met, would simultaneously ensure national goals without leaving anyone behind.

}



\maketitle


\newpage 

\section{Main}\label{main}

The Sustainable Development Goal 2 (SDG 2) sets out to end hunger, achieve food security, improve nutrition, and promote sustainable agriculture by 2030 while promising to ``leave no one behind” \cite{united_nations_transforming_2015}. The importance of smallholders in meeting this goal is prescribed in SDG target 2.3 and the objective to double their agricultural productivity. Despite the existence of proposed indicators, this target has been notoriously difficult to monitor, and notwithstanding ongoing complementary initiatives to close those data gaps, including the 50x2030 initiative \cite{winters_facilitating_2022}, large data gaps persist temporally and spatially \cite{fao_crops_2022}.  Furthermore, even when large area estimates of national progress exist, local disparities make it difficult to evaluate who is benefiting most from national progress and who may continue to be left behind \cite{seery_hitting_2019}. This lack of cost-effective and fine monitoring, in turn, makes it difficult for countries to design policies and support programs that can meet SDG 2.3 and close the gap between producers at the same time.

Satellite-enabled remote monitoring of agronomic output offers significant opportunity to improve monitoring, and yet while a number of demonstrations exist for predicting crop yields on smallholder farms \cite{azzari_understanding_2021, jin_smallholder_2019, burke_satellite-based_2017}, there remains a fundamental need to advance crop monitoring services in smallholder systems across the world \cite{mehrabi_research_2022}, on annual or seasonal basis and at high-resolution.  
 
Here, we apply a new spatially explicit time series of maize cover and productivity for Rwanda \cite{fankhauser_high_2024} at requisite spatial (10 m) and temporal (seasonal; within days post-season) resolution to detect trends across 15 thousand villages, performance between them, and contributions to national trends. The geographic and crop specific focus are convenient, but act as a demonstration that could be carried to other nations and crops. We, then, return to the question of what it would take to bring all villages on track in the country alongside alternative policies to achieve equity. 

\begin{figure}[hbt!]
\centering
\includegraphics[width=0.5\textwidth]{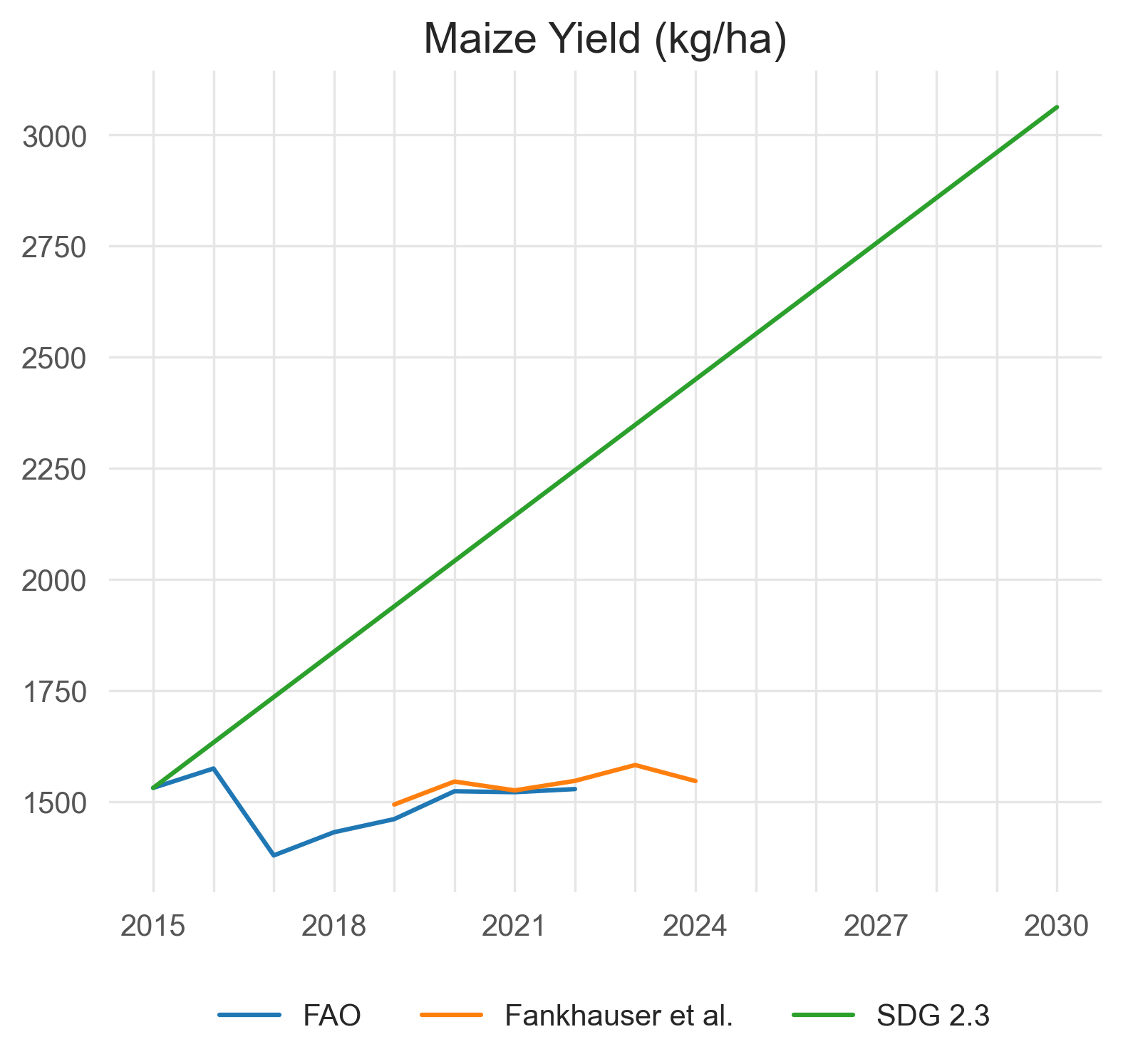}
\caption{Rwanda's national progress towards SDG 2.3 shown through maize yields. The green line demonstrates the linear growth rate required to meet SDG 2.3, the blue line represents national averages published by FAO from 2015-2022 \cite{fao_crops_2022}, and the orange line is the average yield observed from our high-resolution dataset from 2019-2024 \cite{fankhauser_high_2024}.}
\label{fig:sdg}
\end{figure}

Satellite monitoring, which agrees with and extends national survey data \cite{fao_crops_2022}, shows maize yields in Rwanda are considerably below target yields that would realize a doubling in productivity over 2015-2030 (Fig. \ref{fig:sdg}). Under the status quo, Rwanda is not on track to meet SDG 2.3, in spite of national policies and government efforts to bolster productivity \cite{ministry_of_agriculture_and_animal_resources_national_2017}. 

\begin{figure}[hbt!]
\centering
\includegraphics[width=1\textwidth]{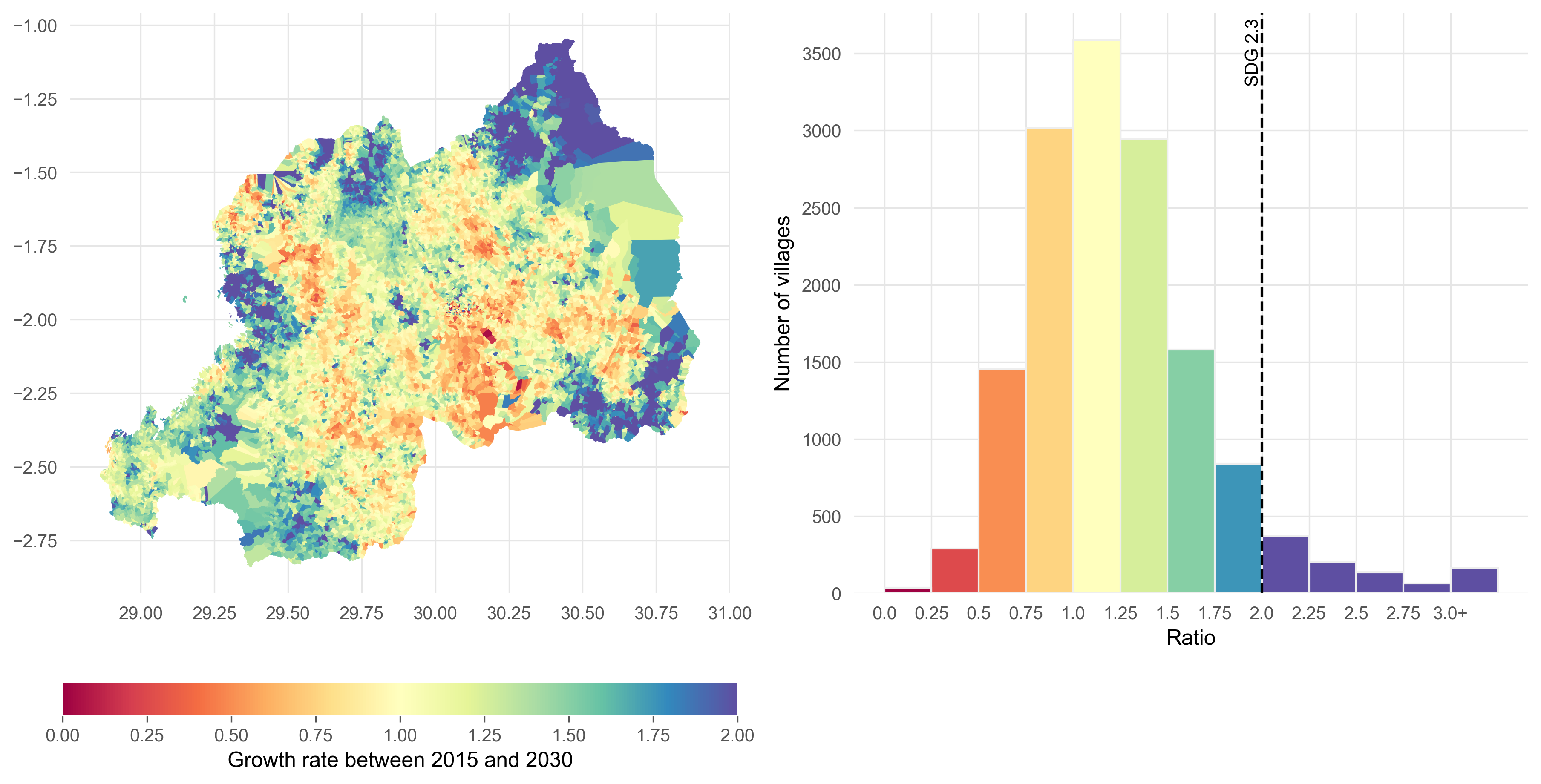}
\caption{Rwandan villages on and off track to meet SDG 2.3. A ratio of 2.0 or higher indicates that maize productivity is projected to double by 2030 based on current growth rates. See Supplementary Figs. \ref{fig:map:lpi} and \ref{fig:map:upi} for conservative and optimistic estimates, respectively.}
\label{fig:map}
\end{figure}

Furthermore, progress to meeting SDG 2.3 is distributed unequally throughout the country (Fig. \ref{fig:map}). Despite demonstrating progress toward roughly one-fifth of the goal nationally, only a small percentage of villages (Table \ref{tbl:scenarios}) will meet SDG 2.3 if the observed linear rate of growth in the prior four years (2019-2023) continues. We find the Eastern Province, particularly in the north and southeast, and central Western Province are home to the majority of villages achieving notable progress in doubling productivity, with many areas in the central part of the country experiencing stagnant, or even negative, growth rates.

We also note large disparities between the lowest and highest yielding producers in the country (Supplementary Fig. \ref{sup:fig:qntls}). Yield gaps between these were declining in the country, but began rising again during the COVID-19 pandemic, increasing inequality. Preliminary 2024 data suggests that this trend is reversing and the productivity gap between villages may be closing again; however, it is a sign of modest progress --- currently the average yield among villages in the 90th percentile was roughly 2.4 times that among villages in the 10th percentile (2166 vs. 915 kg/ha).

\begin{table}[hbt!]
\caption{National and village-level outcomes towards SDG 2.3 to double productivity for maize under potential scenarios}
\label{tbl:scenarios}
\tiny
\begin{tabular}{lccccc}
\toprule
Scenario &
  \begin{tabular}[c]{@{}c@{}}Natl SDG \\ Progress 2030 \\ (\% of goal)\end{tabular} &
  \begin{tabular}[c]{@{}c@{}}Additional years \\ to meet SDG \\ (Natl)\end{tabular} &
  \begin{tabular}[c]{@{}c@{}}Village SDG \\ Progress 2030 \\ (\% of villages)\end{tabular} &
  \begin{tabular}[c]{@{}c@{}}Equality 2030 \\ (Ratio) \end{tabular} &
  \begin{tabular}[c]{@{}c@{}}Greatest growth \\ rate after 2024 \\ (kg/ha/year)\end{tabular} \\
\midrule  
Sc1: Current &
  \begin{tabular}[c]{@{}c@{}}18.6 \\ (9.6 - 23.7)\end{tabular} &
  \begin{tabular}[c]{@{}c@{}}65.6 \\ (inf - 19.3)\end{tabular} &
  \begin{tabular}[c]{@{}c@{}}6.4 \\ (12.1 - 3.4)\end{tabular} &
  \begin{tabular}[c]{@{}c@{}}3.5 \\ (16.6 - 2.3)\end{tabular} &
  \begin{tabular}[c]{@{}c@{}}422 \\ (191 - 826)\end{tabular} \\
Sc2: National SDG &
  100 &
  0 &
  \begin{tabular}[c]{@{}c@{}}51.5 \\ (51.1 - 52.6)\end{tabular} &
  \begin{tabular}[c]{@{}c@{}}1.6 \\ (2.2 - 1.4)\end{tabular} &
  \begin{tabular}[c]{@{}c@{}}212 \\ (107 - 317)\end{tabular} \\
Sc3: Village SDG &
  100 &
  0 &
  100 &
  \begin{tabular}[c]{@{}c@{}}2.5 \\ (5.3 - 2)\end{tabular} &
  \begin{tabular}[c]{@{}c@{}}736 \\ (567 - 1614)\end{tabular} \\
Sc4: Equitable &
  \begin{tabular}[c]{@{}c@{}}84.1\\ (110.5 - 75.2)\end{tabular} &
  \begin{tabular}[c]{@{}c@{}}1.3 \\ (-0.8 - 2)\end{tabular} &
  \begin{tabular}[c]{@{}c@{}}40.8 \\ (54.8 - 25.6)\end{tabular} &
  1.0 &
  \begin{tabular}[c]{@{}c@{}}402 \\ (321 - 434)\end{tabular} \\
\begin{tabular}[c]{@{}c@{}}Sc5: Equitable \\ + Natl SDG\end{tabular} &
  100 &
  0 &
  \begin{tabular}[c]{@{}c@{}}52.0 \\ (50.7 - 52.3)\end{tabular} &
  1.0 &
  \begin{tabular}[c]{@{}c@{}}440 \\ (305 - 513)\end{tabular} \\
\bottomrule
\end{tabular}
\footnotetext{Mean estimate with 95\% prediction interval in parentheses. 

Scenarios relate to business as usual projections (Sc1) or, alternatively, prioritizing progress nationally (Sc2), by village (Sc3), to achieve parity between producers (Sc4), or to gain both parity and national targets (Sc5) by 2030. 

To determine progress towards SDG 2.3 we linearly regressed observed maize yields 2019-2023 by village to get a village-specific average growth rate and, then, used this rate to derive an estimate of yield for the SDG baseline year (2015) and end year (2030). National progress was derived from comparing mean projected yield in 2030 to a doubling of the mean yield at baseline. From the average linear growth rate, we also calculated the number of additional years, if any, needed to meet the goal. For village-level progress, we noted the number of villages succeeding in doubling productivity between the years relative to the total number of villages in the country. Equality was indicated by the ratio in mean yield between the highest and lowest producers (i.e. 90th and 10th percentiles, respectively) in 2030. The greatest growth rate was the maximum annual village-level gain in maize yield required to realize the given scenario; thus, other villages exhibited growth rates less than or equal to this number.}

\end{table}

Under current business as usual projections (Scenario: Sc1; Table \ref{tbl:scenarios}), Rwanda is expected to increase national average maize yield by 2030 to 18.6\% of the goal, but sub-national analysis reveals that only 6.4\% of villages will experience a doubling in productivity and inequality between the highest and lowest yielding producers will remain high. Current growth is inadequate to meet SDG 2.3, now, and perhaps soberingly, even within the next 65 years.

Continued efforts for targeted interventions may produce different, more favorable outcomes (Table \ref{tbl:scenarios} and Supplementary Fig. \ref{sup:fig:scenarios}). For example, a policy to uniformly increase yields (Sc2) in every village by 212 kg/ha/year for the next six years would not only meet SDG 2.3 but also improve equality and bring up over half of the villages. To double productivity by 2030 in every village (Sc3), growth rates would need to be as high as 736 kg/ha/year in some villages, but inequalities would remain. Alternatively, an equity-focused approach, or closing the yield gap in the country by increasing every village's yield to that projected in 2030 for the highest yielding producers (Sc4), would double productivity in 41\% of villages and nearly meet SDG 2.3 nationally at much lower annual growth rates. Finally, to achieve equality while also asserting reaching SDG 2.3 (Sc5), the maximum growth rate across villages would need to increase by only 40 kg/ha/year over that designed to achieve equality alone. The reality of meeting these national and local targets will require drastic interventions, with many factors mediating maize productivity, but there are opportunities to accelerate growth (see Section \ref{feasibility} for sensitivity analyses and discussion). 

Achieving SDG 2 is estimated to cost US\$33 billion year-over-year, with a significant share (58\%) coming from national budgets \cite{laborde_ceres2030_2020}. However, and perhaps most critically, national statistics agencies that collect data to monitor the SDGs are massively underfunded \cite{department_of_economic_and_social_affairs_sustainable_2023}, limiting the availability of key datasets to target and detect policy induced change. Satellite-enabled remote monitoring of agricultural production offers a cost-effective technology for monitoring progress towards SDG 2, and for identifying performance at high spatial resolution. As an example, the pipeline we use here \cite{fankhauser_high_2024} offers open access, remote, wall-to-wall, and near-real time agricultural monitoring for Rwanda for $\sim$USD\$20 per season, monumental benefits that could be scaled to other data sparse countries in sub-Saharan Africa. As such we see much promise for the extension, use, and deployment of these technologies for advancing progress towards SDG 2.3 more widely.

It is important to view our results in context. What data exist shows that many countries are not making meaningful progress towards SDG 2.3 --- and in particular yields are not rising fast enough in sub-Saharan Africa \cite{sachs_implementing_2023}. The Green Revolution that spurred widespread agriculture development between 1960 and 2000 increased maize yields by 157\% in developing countries, but the new technologies were not suited for African agricultural systems and many countries were left behind \cite{pingali_green_2012}. We risk repeating the lack of support for marginal environments and populations that gives rise to inequalities and the shortfalls in progress towards the SDGs demonstrates we are already failing. We need targeted investment in research, infrastructure, finance and market development, and policy for these contexts \cite{pingali_green_2012} and the means to identify the low productivity systems at risk of being left behind. 

Reaching equity and national targets for agricultural systems can be complementary goals, but policymakers have a number of synergies and trade-offs to navigate. First, targeting uniform yield growth across the country, which may be preferred due to ease of enrollment and monitoring, addresses inequality implicitly because all villages, including marginal environments, are treated, but it also preserves the baseline disparity between villages. Second, targeting reductions in inequality alone and directing resources to low and moderate yield producing villages contributes to national progress and has companion benefits: the return on investment of high yielding crop varieties is much greater for low potential areas \cite{fan_returns_2001} and gains in smallholder yields generates greater economic growth and poverty reduction than investments in other sectors or on large farms \cite{pingali_green_2012, larson_rural_2020}. Third, continuing to invest in high yield areas, while devoting most resources to marginal environments, keeps prices low and capitalizes on the spillover effects of stronger rural economies and technology and infrastructure development \cite{pingali_green_2012}, but can neglect a large number of farmers in the country. 

Significant effort is required to develop appropriate agricultural policies. A key constraint  to improving yields is the lack of financing, institutions, and an enabling environment \cite{von_braun_marginality_2014}, factors that have been slow to materialize for many smallholders in sub-Saharan Africa. Researchers can, however, assist in this task by leveraging technologies to help policymakers decide which policies, interventions, and programs may realize the growth rates required and set targets that are needed to realize goals of national and local government. The monitoring system based on high- spatial and temporal resolution data presented here may provide a solid foundation for evidence-based decision making, with future focus being on monitoring other staples and "orphan" crops \cite{pingali_green_2012} as diversification and localization of nutrition and food security become a complementary and much needed focus of agricultural policies. 

With this approach, we are hopeful that continued interplay between scientists, policymakers, and grass roots organizations may assist in identifying, targeting, and helping to achieve agricultural development goals and targets for all in ways and modes that were previously inconceivable.

\section{Methods}\label{methods}

\subsection{Data}\label{data}

We use high-resolution maize yield time series for Rwanda and the  machine learning pipeline from \cite{fankhauser_high_2024}. Briefly, we predicted land use, maize cover, and maize yields at every 10 m pixel in Rwanda for 11 maize-growing agricultural seasons across 2019 - 2024. From 60,000 field-level observations and 9,000 crop cuts for maize, we built gradient boosted tree models to predict maize cover and yields from Sentinel-2 Level-2A optical imagery \cite{esa_sentinel-2_2023} and satellite-derived land-surface temperature \cite{wan_modisterra_2021} and rainfall \cite{funk_climate_2015}. Preliminary predictions are available within days after the end of a season and later can be harmonized with national statistics. When possible, we normalize predictions to national agricultural statistics \cite{nisr_seasonal_2023-1} to agree with existing (coarser) reference data that policymakers use for decision making. During the 2019-2023 seasons, maize cover was classified with 83\% accuracy and maize yields were predicted with a root mean squared error (RMSE) of 700 kg/ha scaled to national annual means. This pipeline includes open access to the data products at \cite{fankhauser_high_2024} and low-cost ($\sim$US\$20)  remote, wall-to-wall, and near-real time agricultural monitoring for Rwanda, that could be scaled to other data sparse countries in sub-Saharan Africa.  

For the present analysis, we averaged yields by village, Rwanda's smallest administrative unit. This aided in interpretation and significantly reduced the impact of residual error at 10 m pixels on downstream analysis. Supplementary Fig. \ref{fig:error} shows that the bootstrapped model error among 1153 pixels (the fewest number of pixels used to calculate means between villages in the lowest and highest yield deciles in Supplementary Fig. \ref{sup:fig:qntls}) was a negligible 20 kg/ha on average and that the mean yield converged after around 300 x 10 m pixel samples, showing how village aggregation smoothed much of the model error. 

In Rwanda, maize is cultivated twice per year, during Season A from September to February and again during Season B from March to June \cite{nisr_seasonal_2023-1}. To make estimates comparable to an annual monitoring cycle we aggregated seasonal village yields by taking the annual mean of maize yields weighted by the amount of area under maize cultivation in each season. In this analysis, we treated 2024 Season A as preliminary because at the time of writing we had not yet observed Season B for the year; thus, values from 2024 do not represent a weighted average as in other years. However, they are still likely indicative because Season A is the primary agricultural season \cite{nisr_seasonal_2023-1} and yields were generally more heavily weighted towards it in past years. An interesting follow-on analysis would be using the higher resolution seasonal data to see how smallholders in Rwanda are adapting to changes in seasonal growing conditions. 

\subsection{Analysis}\label{analysis}

To determine current national progress towards SDG 2.3 we took the average maize yield observed in our high-resolution dataset among all villages in each year 2019-2024 and compared it the linear rate of change required to double the national baseline yield measured by FAO in 2015 \cite{fao_crops_2022}. 

We took the average observed maize yield among villages in each yield decile cohort to measure equality between producers of different levels. We defined cohorts in 2019 and followed them through time until present 2024. Equality was indicated by the ratio in mean yield between the highest and lowest producers (i.e. 90th and 10th percentiles, respectively). Upper and lower bounds for the ratio were generated from taking the 2.5th and 97.5th percentile of pairwise comparisons between every village in either group. 

For village-level projections of progress, we linearly regressed observed maize yields 2019-2023 (and in Supplement \ref{sup:y24}, 2019-2024) by village to get a village-specific average growth rate and, then, used this rate to derive an estimate of yield for the SDG baseline year (2015) and end year (2030). If the ratio between predicted yields in 2030 and 2015 was equal to 2 or greater than that village was said to be on track to meet SDG 2.3. Other studies have shown maize and cereal yields to change linearly and have used linear regressions to project yields in other years \cite{hafner_trends_2003, ray_yield_2013, jaggard_possible_2010, fischer_breeding_2010}. The mean linear prediction is presented in the main text, but uncertainty in the regression was characterized by upper and lower prediction intervals and so represent conservative and optimistic runs of the analysis which we present in Supplement \ref{sup:pis}. 

\subsection{Technical feasibility}\label{feasibility}

The reality of meeting national and local targets will require drastic policy interventions, given observed trends over the last 5 years (Fig. \ref{fig:sdg}). Moreover, biophysical limitations can present yield ceilings with technical challenges to overcome. When projected yields are capped to the highest current observed maize yield between 2019-2023 in each respective agro-ecological zone (AEZ), we find Rwanda is unable to meet SDG 2.3 (Supplementary Table \ref{tbl:aez_scenarios}). This is because maximum observed yields within the country (1582 - 2913 kg/ha) are too low to allow for the growth necessary to double productivity in the country (the national target is 3063 kg/ha). However, studies have found maize yields increase to 4000 kg/ha in Rwanda with better pest management \cite{silvestri_analysing_2019} and fertilizer application \cite{bucagu_determining_2020}, and yields in Rwanda are significantly less than that demonstrated by industrial agriculture in high-income countries: the United States produces maize with efficiencies of 10,880 kg/ha on average \cite{fao_crops_2022}. In fact, some researchers have estimated the yield potential for cereals, including maize, is over 10,000 kg/ha in Rwanda, with suitable inputs \cite{tian_will_2021}. Raising yields in this manner will require significant investments, but may allow for more optimistic achievements than those presented by our agro-ecologically constrained yield gap analysis. Investigating the drivers of annual changes in growth may reveal feasible options to raise yields within short time periods. For example, were villages to realize their maximum proven achievable growth rate year-over-year (Sc7; Supplementary Table \ref{tbl:scenarios:add}), many (46\%) would meet SDG 2.3 by 2030, and enough would increase yields in order to meet the SDG nationally. A better understanding of the changes in suitability, both biophysically and socially, in those years may enable improved targeting of agronomic programs.

\backmatter

\bmhead{Supplementary information}

This article has accompanying supplementary files. Supplementary tables and figures are enclosed as an appendix. The high-resolution time series of maize yield in Rwanda can be downloaded at \url{https://doi.org/10.5281/zenodo.10659095} \cite{fankhauser_high_2024}. The data and Python scripts used to produce the present analysis will be provided in the supplementary material upon publication. 

\bmhead{Acknowledgements}

This publication was made possible through support provided by the United States Agency for International Development under the terms of award number 7200AA20FA00021 and the Wellspring Philanthropic Fund. The opinions expressed herein are those of the author(s) and do not necessarily reflect the views of these institutions.

\section*{Declarations}

\begin{itemize}
\setlength\itemsep{1em}

\item Funding

This publication was made possible through support provided by the United States Agency for International Development under the terms of award number 7200AA20FA00021 and the Wellspring Philanthropic Fund. 

\item Competing interests 

The authors declare no competing interests.

\item Ethics approval and consent to participate

Not applicable

\item Consent for publication

Not applicable 

\item Data availability 

The scripts, data, and usage notes to access the high-resolution maize cover and yield time series for Rwanda can be found at the following repository \url{https://doi.org/10.5281/zenodo.10659095} \cite{fankhauser_high_2024}. Village-level aggregated maize yields used in this analysis will be provided in the supplementary material. 

\item Materials availability

Not applicable 

\item Code availability 

The Python code used to generate these analyses will be included in the supplementary material. 

\item Author contribution

Conceptualization: K.F., E.T. and Z.M.; Data curation: K.F.; Formal analysis: K.F.; Funding acquisition: E.T.; Investigation: K.F. and Z.M.; Methodology: K.F. and Z.M.; Project administration: K.F., E.T. and Z.M.; Resources: E.T.; Software: K.F.; Supervision: E.T. and Z.M.; Validation: K.F.; Visualization: K.F. and Z.M.; Writing – original draft: K.F. and Z.M.; Writing - review \& editing: K.F., E.T. and Z.M.

\end{itemize}

\noindent

\newpage

\begin{appendices}

\section{Supplementary Information}\label{sup}

\subsection{Yield inequalities among maize producers}

\begin{figure}[h]
\centering
\includegraphics[width=0.9\textwidth]{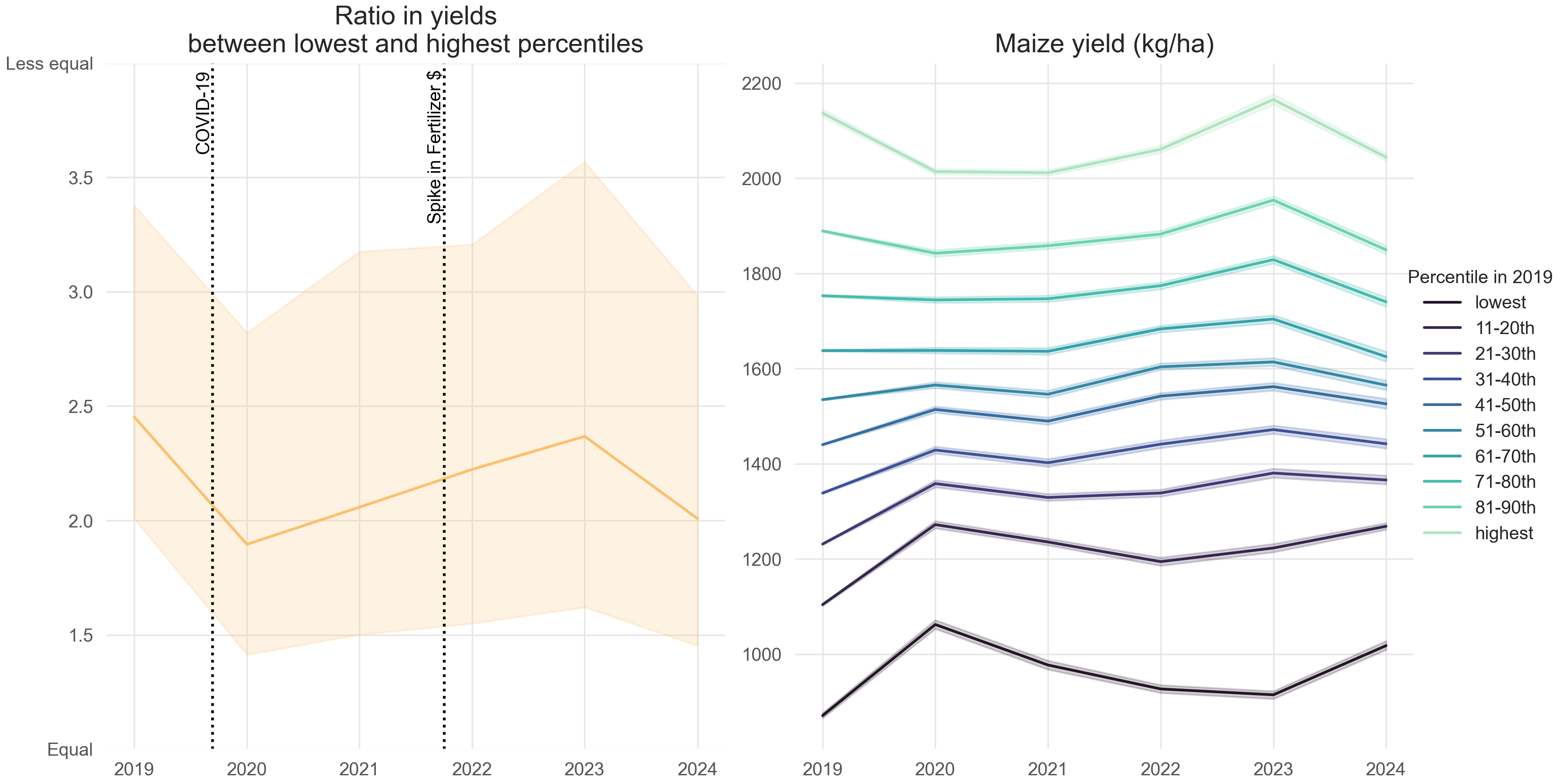}
\caption{Sub-national inequality in maize yields. Average maize yields among villages in each yield decile from 2019 to 2024 (preliminary) and the inequality ratio between the highest and lowest yielding producing villages.}\label{sup:fig:qntls}
\end{figure}

\subsection{Projected outcomes given varying alignment of national and local targets}

\begin{figure}[h]
\centering
\includegraphics[width=1\textwidth]{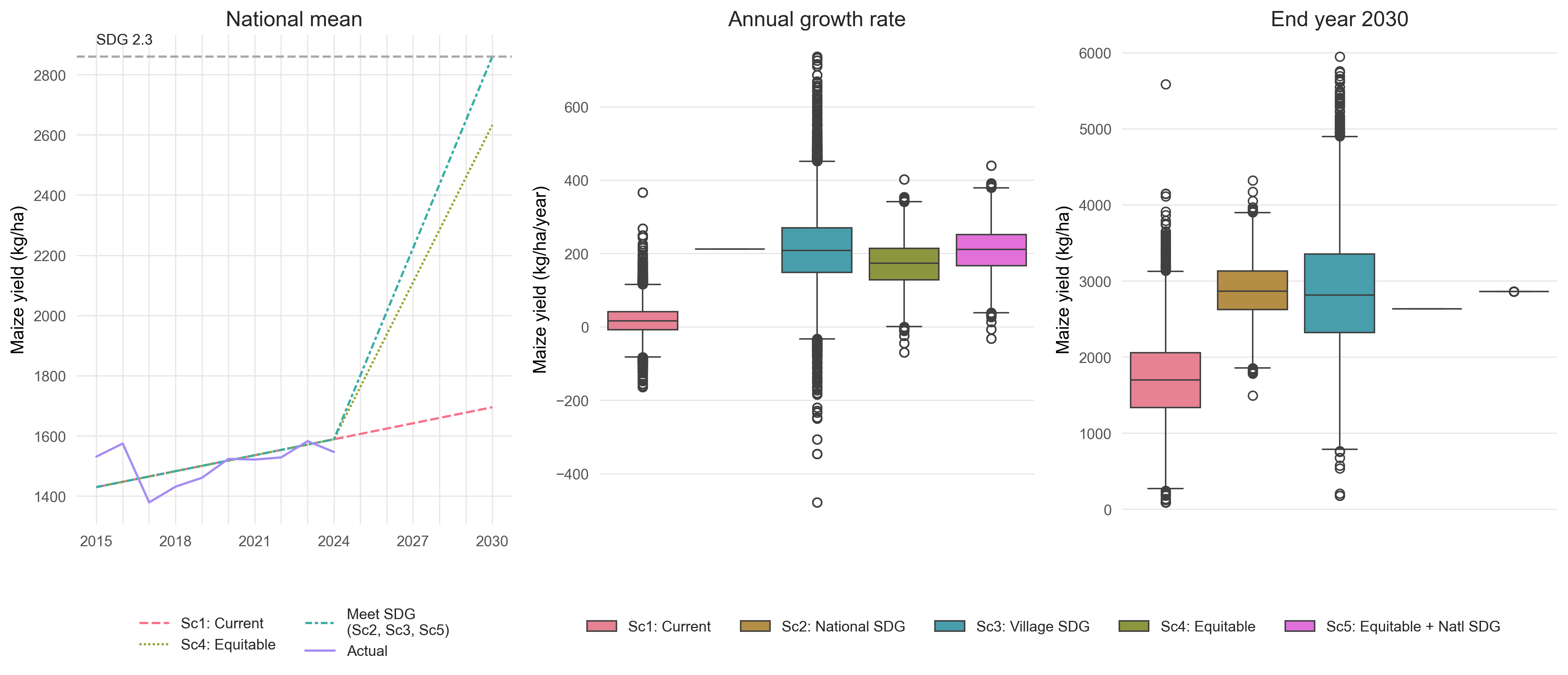}
\caption{Projected yields over the SDG period for each development scenario with requisite growth rates and 2030 yields. See Supplementary Figs. \ref{fig:scenarios:lpi} and \ref{fig:scenarios:upi} for conservative and optimistic estimates, respectively.}\label{sup:fig:scenarios}
\end{figure}

\subsection{Lower and upper prediction intervals for regressed maize yields in year 2030}\label{sup:pis}

\begin{figure}[h]
\centering
\includegraphics[width=1\textwidth]{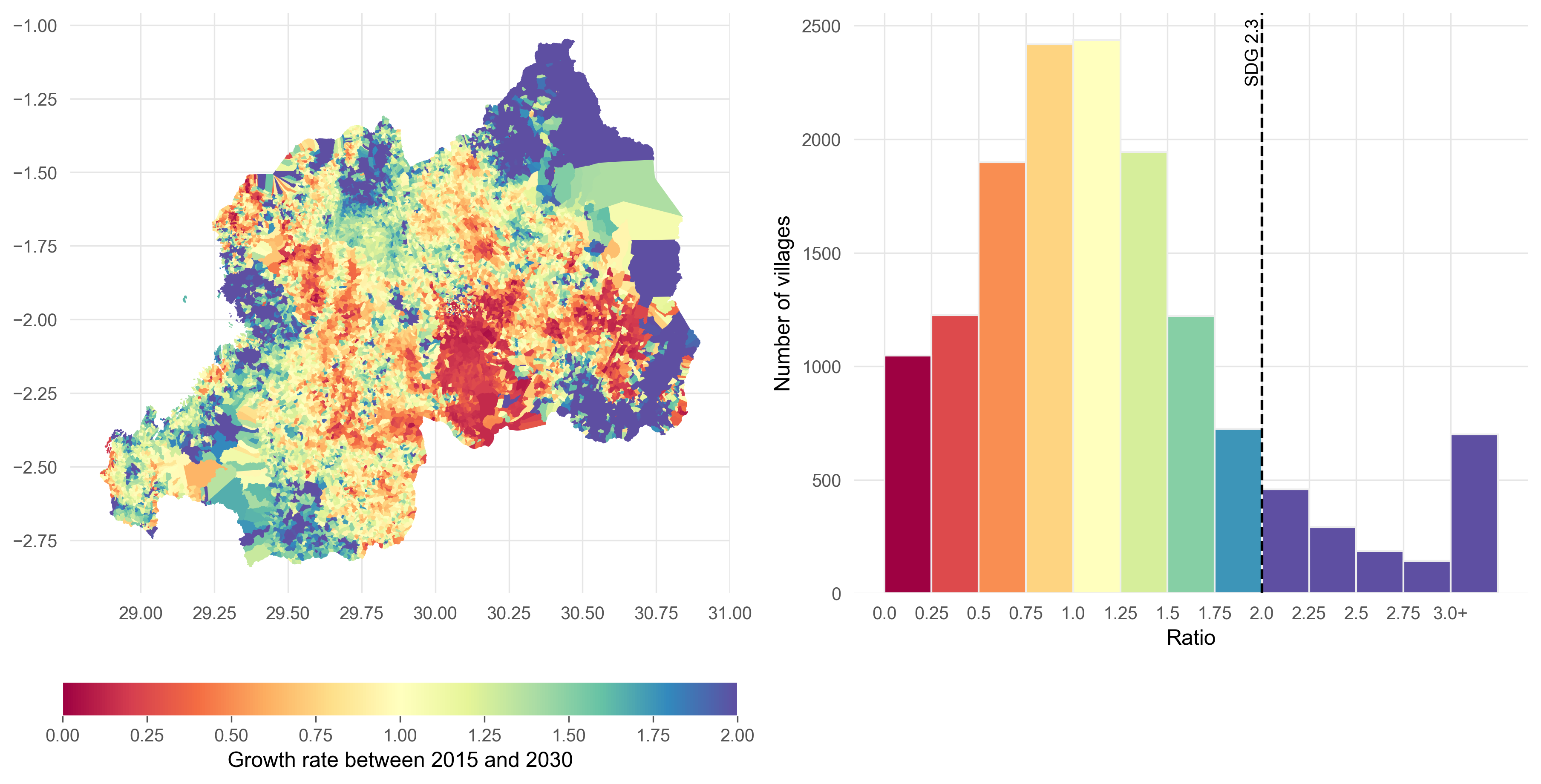}
\caption{Lower prediction interval of Fig. \ref{fig:map}}
\label{fig:map:lpi}
\end{figure}

\begin{figure}[h]
\centering
\includegraphics[width=1\textwidth]{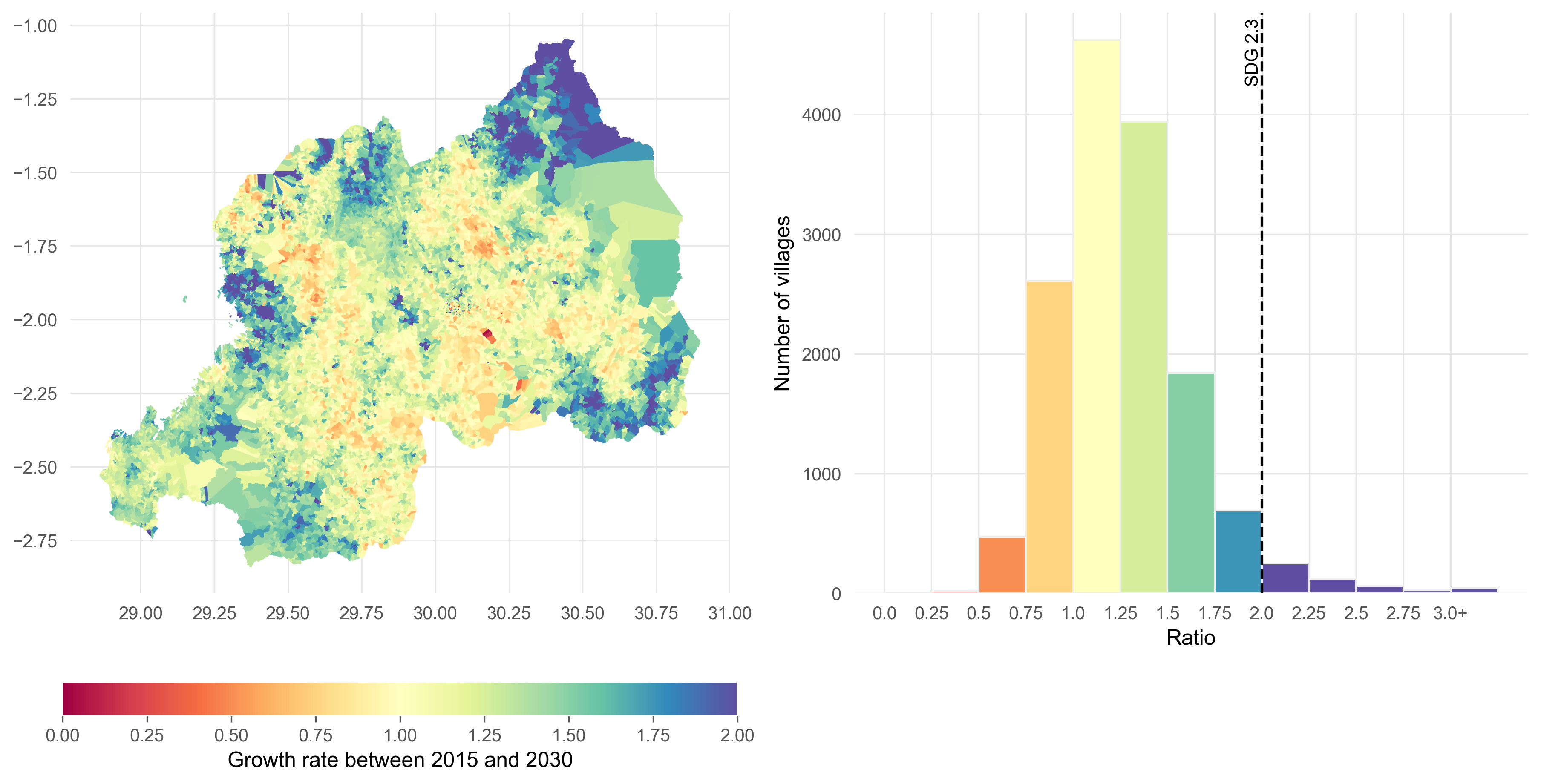}
\caption{Upper prediction interval of Fig. \ref{fig:map}}
\label{fig:map:upi}
\end{figure}

\begin{figure}[h]
\centering
\includegraphics[width=1\textwidth]{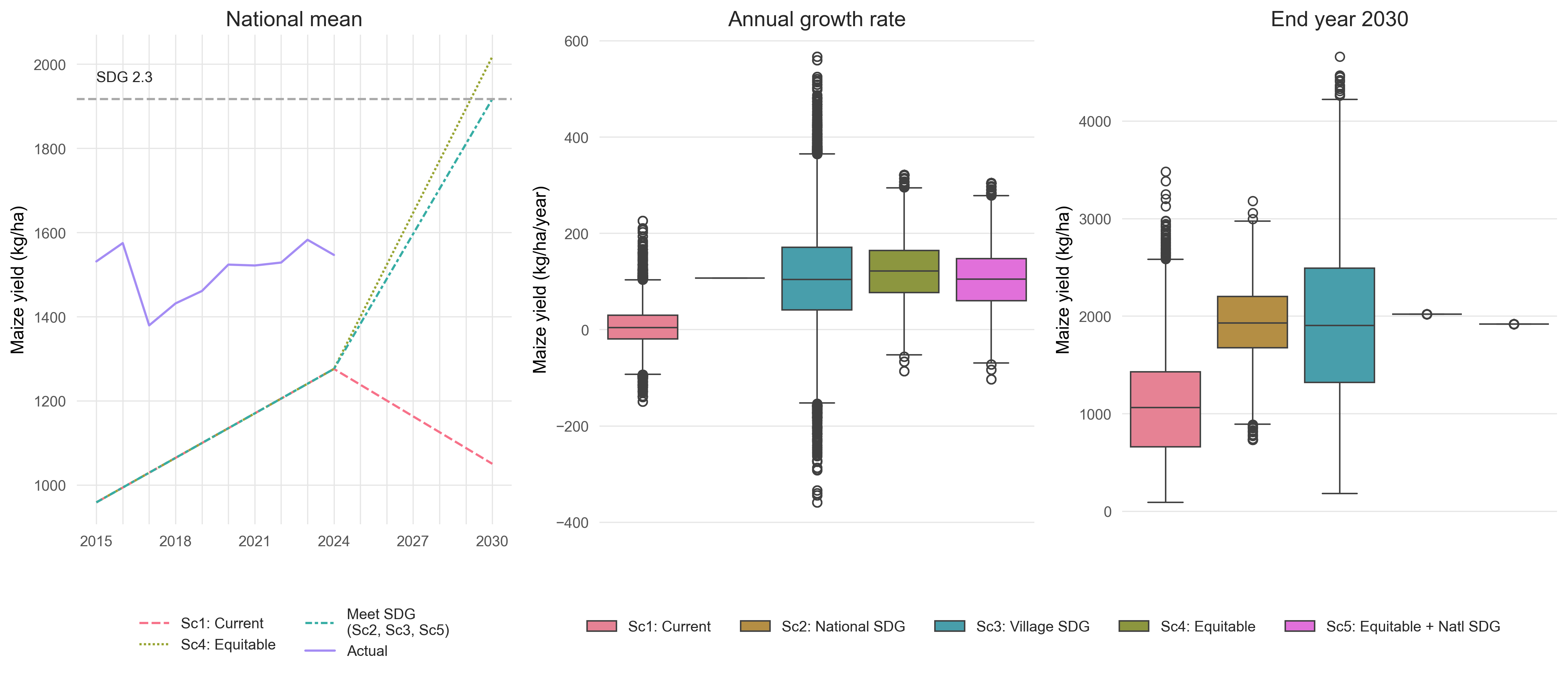}
\caption{Lower prediction interval of Fig. \ref{sup:fig:scenarios}}
\label{fig:scenarios:lpi}
\end{figure}

\begin{figure}[h]
\centering
\includegraphics[width=1\textwidth]{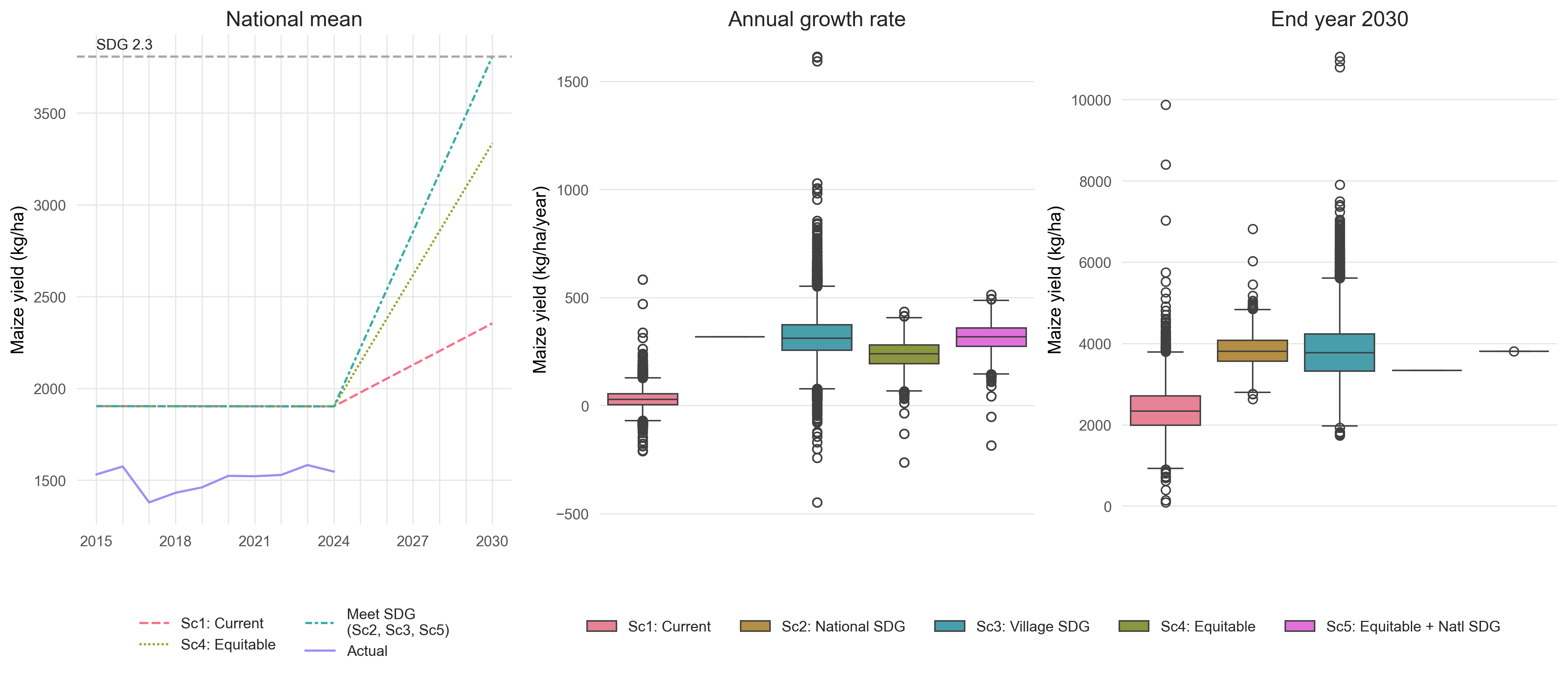}
\caption{Upper prediction interval of Fig. \ref{sup:fig:scenarios}}
\label{fig:scenarios:upi}
\end{figure}

\clearpage

\subsection{Results with preliminary data from year 2024}\label{sup:y24}

\begin{figure}[h]
\centering
\includegraphics[width=1\textwidth]{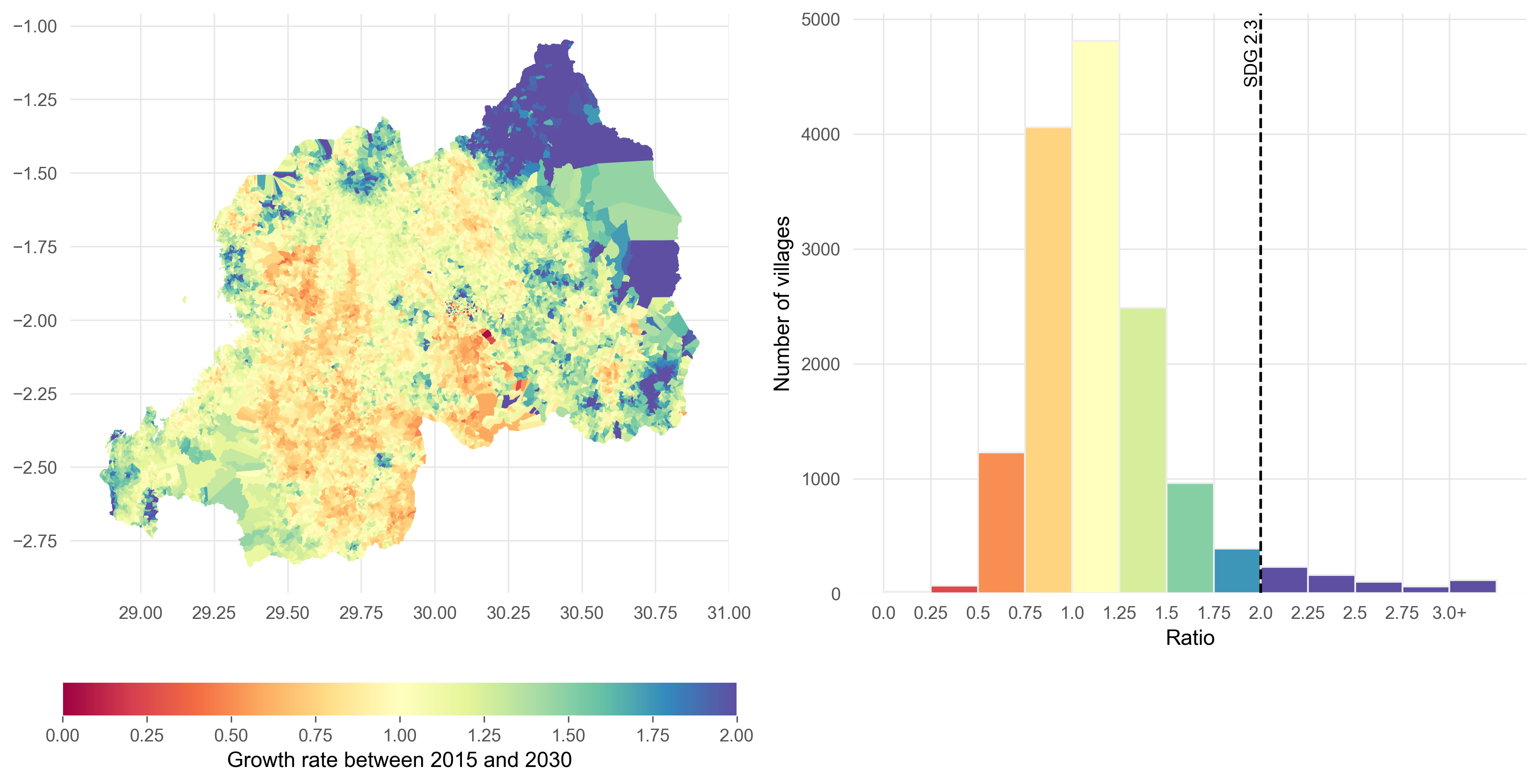}
\caption{Village-level maize yields in 2030 when observed data includes preliminary data from year 2024, companion to Fig. \ref{fig:map}}
\label{fig:map:prelim}
\end{figure}

\begin{figure}[h]
\centering
\includegraphics[width=1\textwidth]{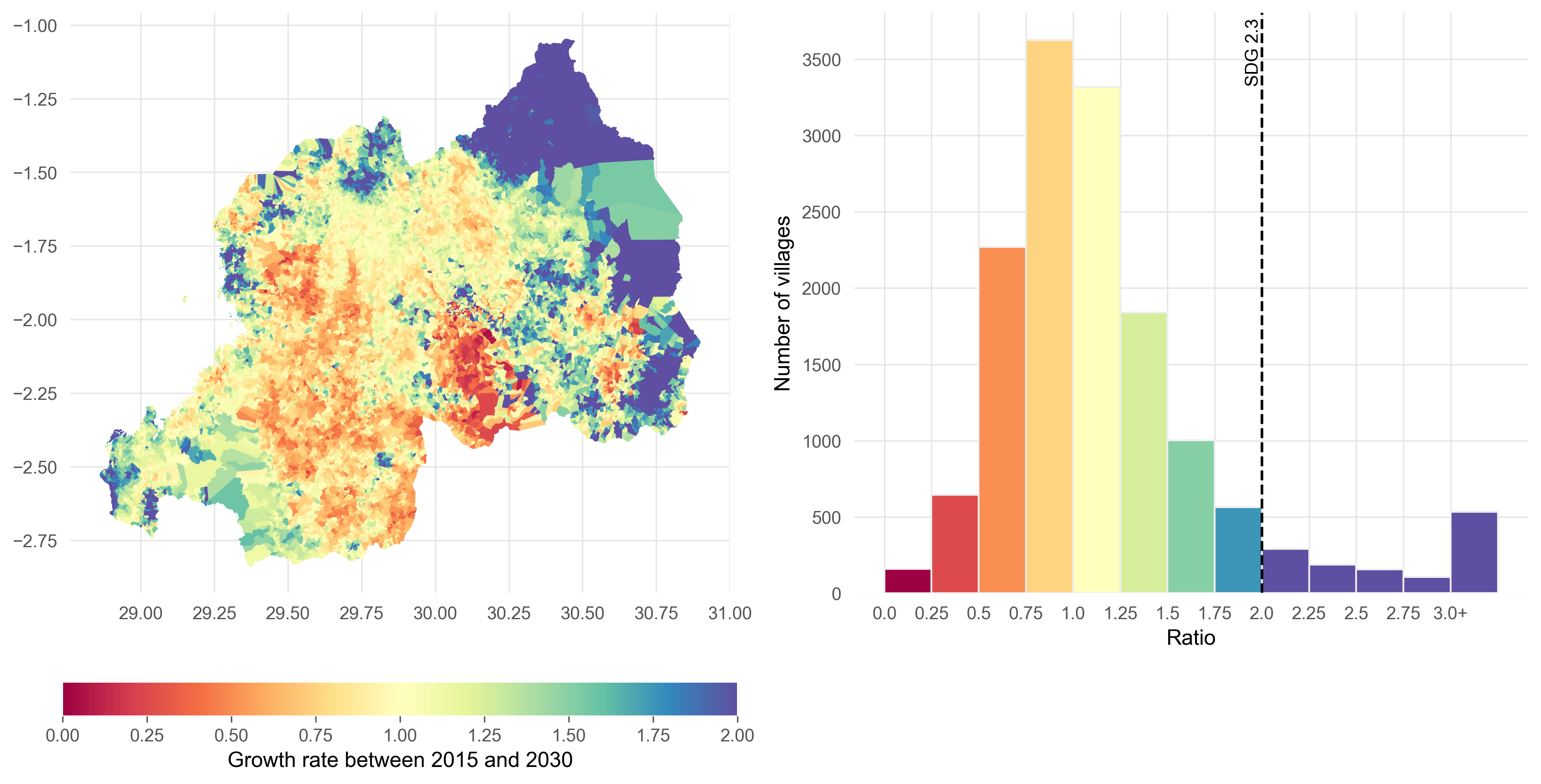}
\caption{Lower prediction interval of Fig. \ref{fig:map:prelim}}
\end{figure}

\begin{figure}[h]
\centering
\includegraphics[width=1\textwidth]{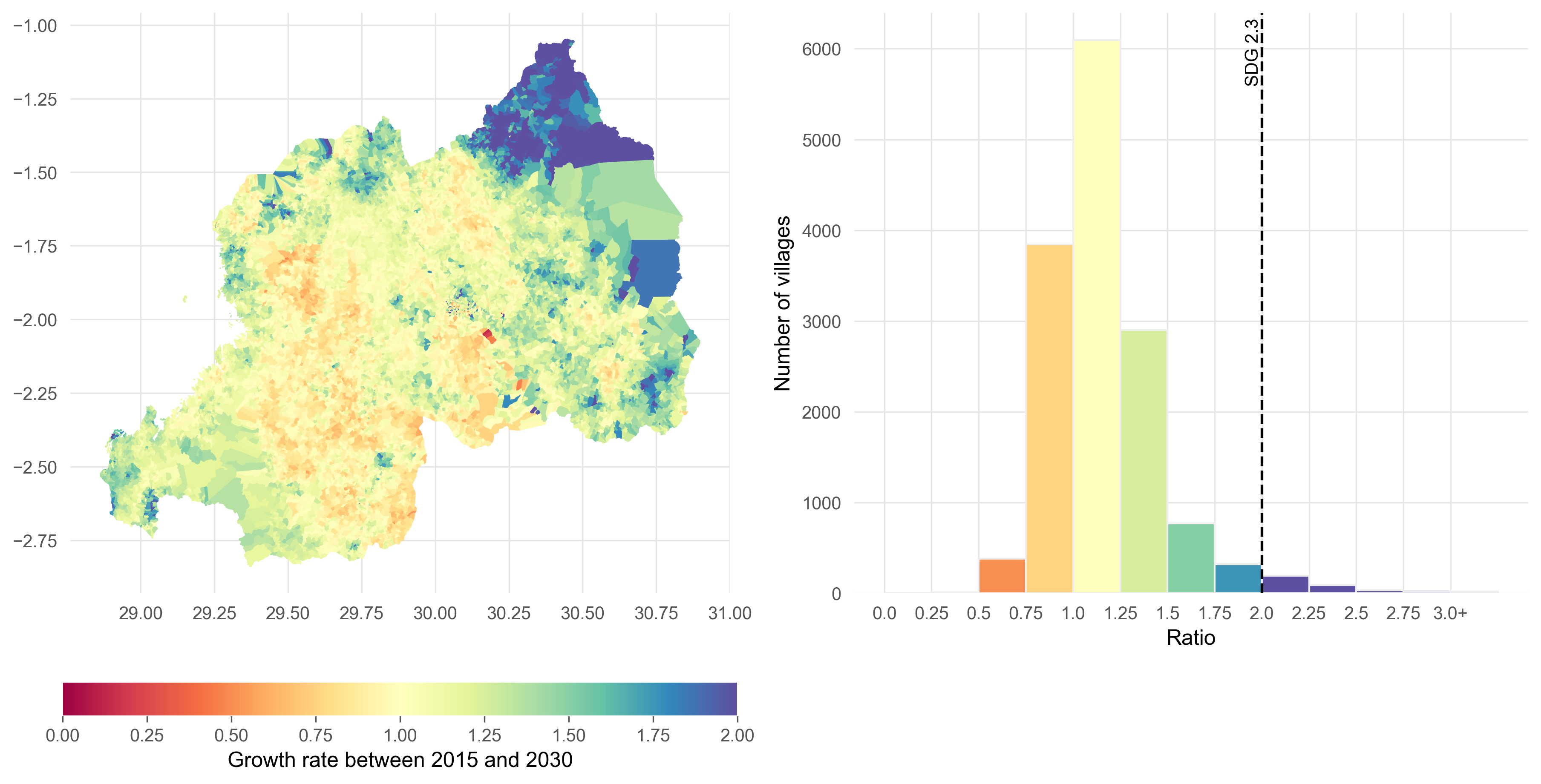}
\caption{Upper prediction interval of Fig. \ref{fig:map:prelim}}
\end{figure}

\begin{table}[h]
\caption{National and village-level outcomes for potential scenarios when observed data includes preliminary data from year 2024, companion to Table \ref{tbl:scenarios}}
\begin{tabular}{lccccc}
\toprule
Scenario &
  \begin{tabular}[c]{@{}c@{}}Natl SDG \\ Progress 2030 \\ (\% of goal)\end{tabular} &
  \begin{tabular}[c]{@{}c@{}}Additional years \\ to meet SDG \\ (Natl)\end{tabular} &
  \begin{tabular}[c]{@{}c@{}}Village SDG \\ Progress 2030 \\ (\% of villages)\end{tabular} &
  \begin{tabular}[c]{@{}c@{}}Equality 2030 \\ (Ratio)\end{tabular} &
  \begin{tabular}[c]{@{}c@{}}Greatest growth \\ rate after 2024 \\ (kg/ha/year)\end{tabular} \\
\midrule
Sc1: Current &
  \begin{tabular}[c]{@{}c@{}}11.3 \\ (7.0 - 13.7)\end{tabular} &
  \begin{tabular}[c]{@{}c@{}}117.3 \\ (inf - 31.6)\end{tabular} &
  \begin{tabular}[c]{@{}c@{}}4.6 \\ (8.7 - 2.5)\end{tabular} &
  \begin{tabular}[c]{@{}c@{}}2.7 \\ (5.2 - 2.1)\end{tabular} &
  \begin{tabular}[c]{@{}c@{}}422 \\ (141 - 824)\end{tabular} \\
Sc2: National SDG &
  100 &
  0 &
  \begin{tabular}[c]{@{}c@{}}49.7 \\ (49.3 - 50.1)\end{tabular} &
  \begin{tabular}[c]{@{}c@{}}1.5 \\ (1.8 - 1.4)\end{tabular} &
  \begin{tabular}[c]{@{}c@{}}228 \\ (131 - 325)\end{tabular} \\
Sc3: Village SDG &
  100 &
  0 &
  100 &
  \begin{tabular}[c]{@{}c@{}}2.5 \\ (4.4 - 2)\end{tabular} &
  \begin{tabular}[c]{@{}c@{}}716 \\ (485 - 1540)\end{tabular} \\
Sc4: Equitable &
  \begin{tabular}[c]{@{}c@{}}65.9 \\ (84.5 - 57.8)\end{tabular} &
  \begin{tabular}[c]{@{}c@{}}3.5 \\ (1.5 - 4.2)\end{tabular} &
  \begin{tabular}[c]{@{}c@{}}28.6 \\ (42.9 - 15.2)\end{tabular} &
  1.0 &
  \begin{tabular}[c]{@{}c@{}}368 \\ (303 - 380)\end{tabular} \\
\begin{tabular}[c]{@{}c@{}}Sc5: Equitable \\ + Natl SDG\end{tabular} &
  100 &
  0 &
  \begin{tabular}[c]{@{}c@{}}49.5 \\ (48.7 - 50.5)\end{tabular} &
  1.0 &
  \begin{tabular}[c]{@{}c@{}}451 \\ (330 - 514)\end{tabular} \\

\bottomrule
\end{tabular}
\footnotetext{Mean estimate with 95\% prediction interval in parentheses.}
\end{table}

\clearpage

\subsection{Additional scenarios}\label{sup:scenarios}

\begin{table}[h]
\caption{National and village-level outcomes for additional potential scenarios, companion to Table \ref{tbl:scenarios}}
\label{tbl:scenarios:add}
\begin{tabular}{lccccc}
\toprule
Scenario &
  \begin{tabular}[c]{@{}c@{}}Natl SDG \\ Progress 2030 \\ (\% of goal)\end{tabular} &
  \begin{tabular}[c]{@{}c@{}}Additional years \\ to meet SDG \\ (Natl)\end{tabular} &
  \begin{tabular}[c]{@{}c@{}}Village SDG \\ Progress 2030 \\ (\% of villages)\end{tabular} &
  \begin{tabular}[c]{@{}c@{}}Equality 2030 \\ (Ratio) \end{tabular} &
  \begin{tabular}[c]{@{}c@{}}Greatest growth \\ rate after 2024 \\ (kg/ha/year)\end{tabular} \\
\midrule  
\begin{tabular}[c]{@{}c@{}}Sc6: Equitable \\ + Village SDG\end{tabular} &
  \begin{tabular}[c]{@{}c@{}}315.9\\ (386.1 - 480.8)\end{tabular} &
  \begin{tabular}[c]{@{}c@{}}-4.3 \\ (-4.9 - -4.8)\end{tabular} &
  100 &
  1.0 &
    \begin{tabular}[c]{@{}c@{}}954 \\ (762 - 1721)\end{tabular} \\
\begin{tabular}[c]{@{}c@{}}Sc7: Max Achieved \\ Growth Rate\end{tabular} &
  \begin{tabular}[c]{@{}c@{}}96.5\\ (160.3 - 64.1)\end{tabular} &
  \begin{tabular}[c]{@{}c@{}}0.2 \\ (-2.8 - 3.4)\end{tabular} &
  \begin{tabular}[c]{@{}c@{}}46.0 \\ (72.6 - 19.5)\end{tabular} &
  \begin{tabular}[c]{@{}c@{}}2.5 \\ (2.8 - 2.3)\end{tabular} &
  803 \\
\bottomrule
\end{tabular}
\footnotetext{Mean estimate with 95\% prediction interval in parentheses.}
\end{table}

\clearpage

\subsection{Potential yield ceilings by agro-ecological zone}\label{sup:aez}

\begin{figure}[h]
\centering
\includegraphics[width=1\textwidth]{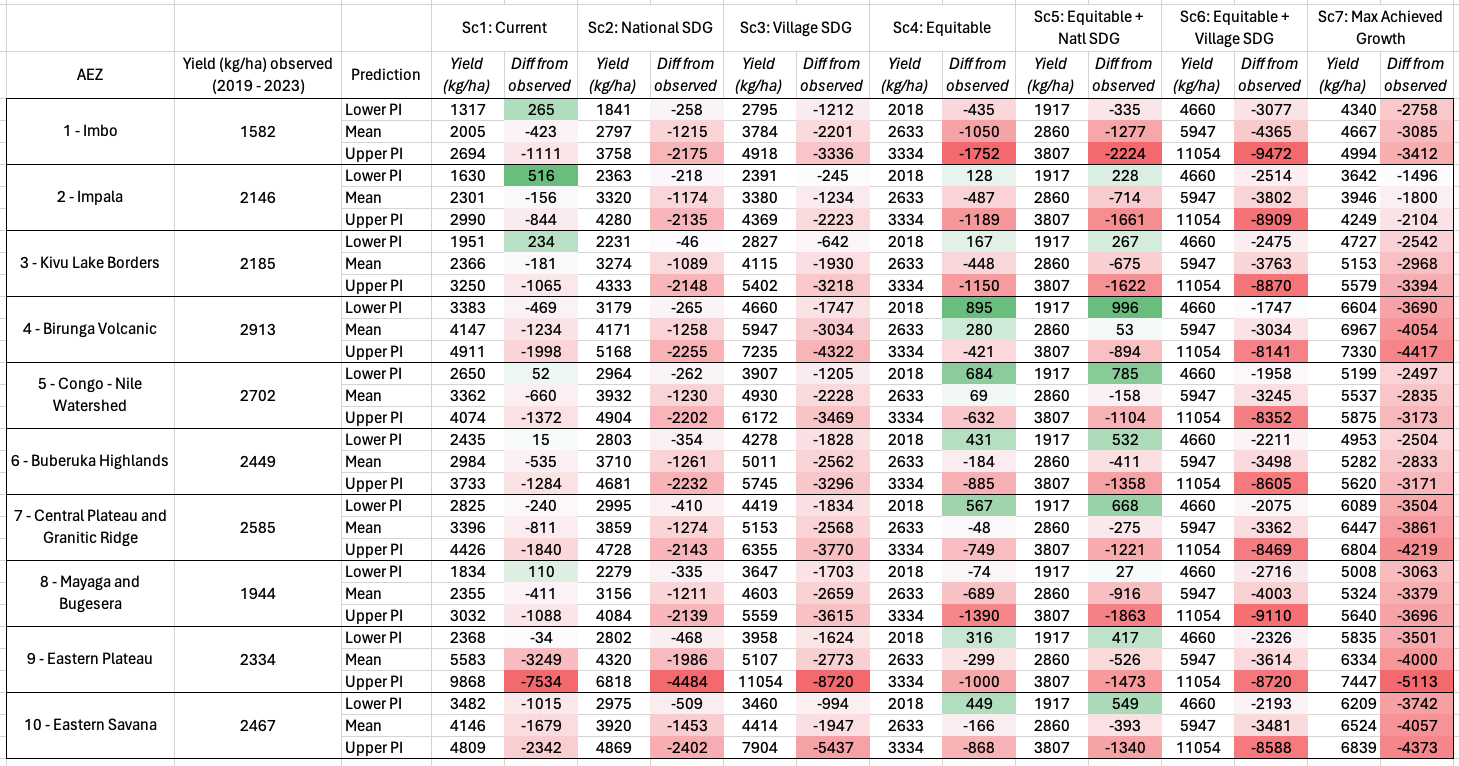}
\caption{Maximum maize yields observed and expected by scenario for each agro-ecological zone (AEZ). The green-to-red scale indicates where projected yields are less than to greater than the hypothetical yield ceiling in each zone.}
\label{fig:tbl:aez}
\end{figure}

\begin{table}[h]
\caption{National and village-level outcomes for potential scenarios when yield is censored to the maximum yield observed 2019-2023 in each village's respective agro-ecological zone, companion to Table \ref{tbl:scenarios}}
\label{tbl:aez_scenarios}
\begin{tabular}{@{}cccccc@{}}
\toprule
Scenario &
  \begin{tabular}[c]{@{}c@{}}Natl SDG \\ Progress 2030 \\ (\% of goal)\end{tabular} &
  \begin{tabular}[c]{@{}c@{}}Additional years \\ to meet SDG \\ (Natl)\end{tabular} &
  \begin{tabular}[c]{@{}c@{}}Village SDG \\ Progress 2030 \\ (\% of villages)\end{tabular} &
  \begin{tabular}[c]{@{}c@{}}Equality 2030 \\ (Ratio)\end{tabular} &
  \begin{tabular}[c]{@{}c@{}}Greatest growth \\ rate after 2024 \\ (kg/ha/year)\end{tabular} \\
\midrule
Sc1: Current      & 17.7 & inf & 6.1  & 3.3 & 187 \\
Sc2: National SDG & 70.5 & inf & 25.4 & 1.5 & 212 \\
Sc3: Village SDG  & 65.1 & inf & 26.2 & 1.8 & 352 \\
Sc4: Equitable    & 10.7 & inf & 2.5  & 1.0 & 227 \\
\bottomrule
\end{tabular}
\footnotetext{Outcomes shown only for mean estimates.}
\end{table}

\clearpage

\subsection{Bootstrapped residual error and convergence to the mean with aggregation}\label{sup:error}

\begin{figure}[h]
\centering
\includegraphics[width=1\textwidth]{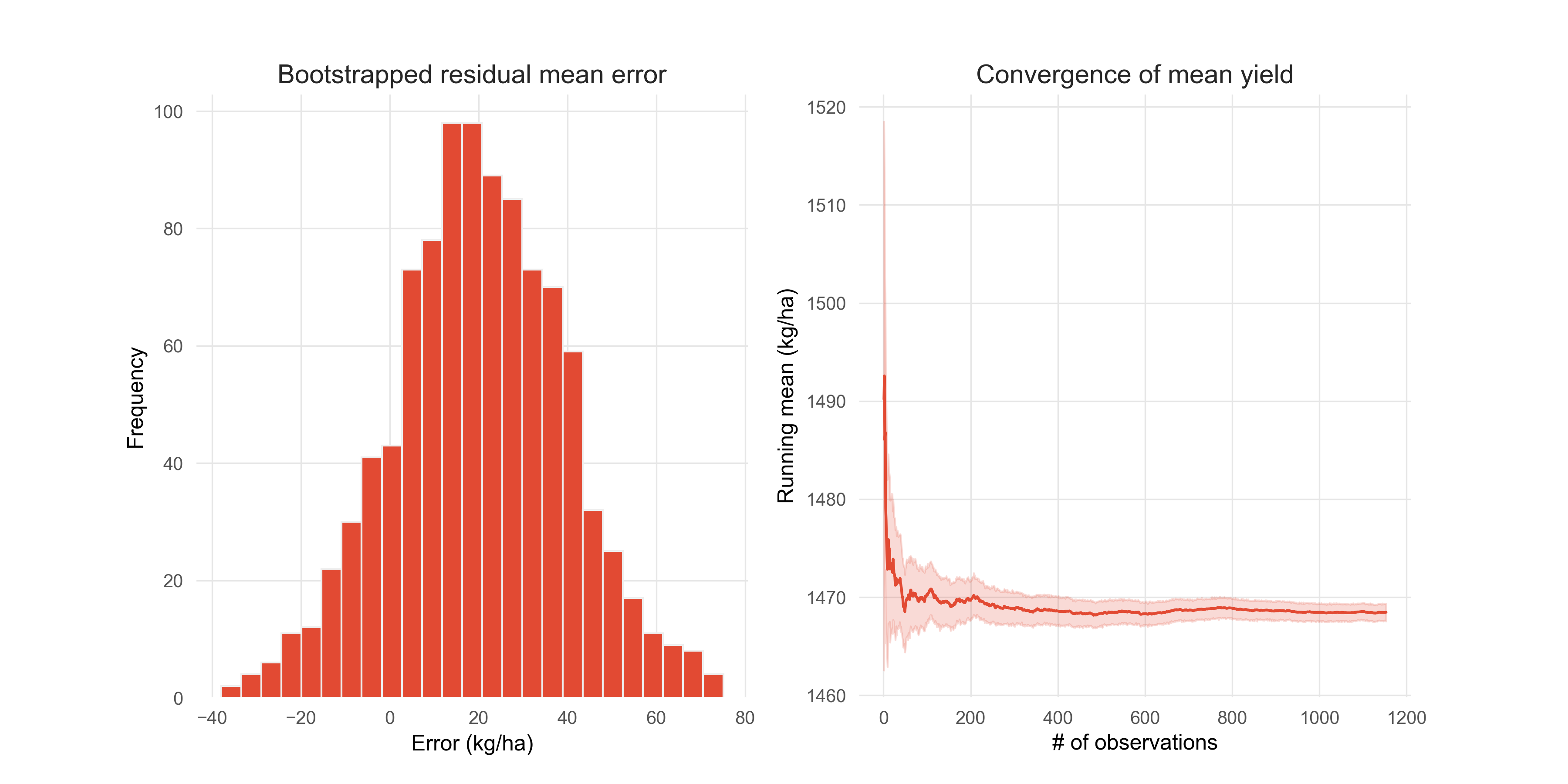}
\caption{Bootstrapped error and convergence of the mean from 1153 samples of residual model error in a test set. Average bootstrapped error was around 20 kg/ha and the mean yield converged after around 300 samples, making the case for village aggregation.}
\label{fig:error}
\end{figure}

\newpage




\end{appendices}


\bibliography{bibliography} 


\end{document}